\documentclass[aps,pra,twocolumn,tightenlines,showkeys,showpacs,groupedaddress]{revtex4}
\usepackage{graphicx}
\usepackage{amsmath}
\usepackage{longtable}

\begin{document}

\title{Convergence of the partial wave expansion of the He ground state}
\author{M.W.J.Bromley}
\email{mbromley@physics.sdsu.edu}
\affiliation{Department of Physics, San Diego State University, San Diego CA 92182, USA}
\author{J.Mitroy}
\email{jxm107@rsphysse.anu.edu.au}
\affiliation{Faculty of Technology, Charles Darwin University, Darwin NT 0909, Australia}

\date{\today}

\begin{abstract}

The Configuration Interaction (CI) method using a very large Laguerre 
orbital basis is applied to the calculation of the He ground state.  
The largest calculations included a minimum of 35 radial orbitals for 
each $\ell$ ranging from 0 to 12 resulting in basis sets
in excess of 400 orbitals.  The convergence of the energy and 
electron-electron $\delta$-function with respect to $J$ (the maximum 
angular momenta of the orbitals included in the CI expansion) were 
investigated in detail.  Extrapolations to the limit of infinite in
angular momentum using expansions of the type 
$\Delta X_J = A_X(J+{\scriptstyle \frac{1}{2}})^{-p} + B_X(J+{\scriptstyle \frac{1}{2}})^{-p-1} + \ldots $,
gave an energy accurate to $10^{-7}$ Hartree and a value of $\langle \delta \rangle$ 
accurate to about 0.5$\%$.  Improved estimates of $\langle E \rangle$ and 
$\langle \delta \rangle$, accurate to 10$^{-8}$ Hartree and 0.01$\%$ 
respectively, were obtained when extrapolations to an infinite radial basis
were done prior to the determination of the $J \to \infty$ limit.   
Round-off errors were the main impediment to achieving even higher 
precision since determination of the radial and angular limits required 
the manipulation of very small energy and $\langle \delta \rangle$ 
differences. 

\end{abstract}


\maketitle

\section{Introduction}

Large configuration interaction (CI) calculations 
of the helium ground state are performed here in order to more precisely 
elucidate the convergence properties of the CI
expansion for this atom.  The general properties of the CI expansion 
have been known since the seminal work of Schwartz \cite{schwartz62a},
which provided the underlying foundation for the later analytic and
computational investigations 
\cite{carroll79a,hill85a,kutzelnigg92a,decleva95a,jitrik97a,ottschofski97a,sims02a}.  
The CI expansion using single center orbitals is slowly convergent with respect
to $J$, the maximum angular momentum of any orbital included in the 
CI expansion.  In particular, the leading term to the energy increment 
is expected to behave at high $J$ as:
\begin{equation} 
\Delta E^{J}  =  \langle E \rangle^J - \langle E \rangle^{J-1} \approx \frac{A_E}{(J+{\scriptstyle \frac{1}{2}})^4}  \label{pE1} \; .  
\end{equation} 

Although there have been a number of very large CI calculations 
performed on helium, all of the earlier calculations using analytic
basis sets treat the 
higher $J$ contributions to the energy with less precision than 
the low $J$ contributions.  Typically, the number of radial
orbitals for the high $\ell$ are smaller than the number of 
low $\ell$.  The justification for this is that the high $\ell$
partial waves make a smaller contribution to the energy and other 
expectation values of than the low $\ell$ orbitals.  At first sight 
this approach would seem reasonable for obtaining estimates of the 
total energy.  However, this approach does lead to problems when 
studying the convergence properties of CI expansion itself.  Here 
it is necessary to ensure that the successive contributions to the
energy are obtained with the same {\em relative} accuracy and this
can hardly be guaranteed with a radial basis that decreases in
size as $\ell$ increases.  Indeed, the evidence suggests that
the dimension of the radial basis should be increased as $J$       
increases if the relative accuracy of the energy is to be
maintained \cite{kutzelnigg99a,mitroy06a}.   

The convergence problems present in CI calculations of atomic
and molecular structure are also present in a much severe 
manner in CI calculations of the positron-atom problem.  The CI 
method has recently been applied to the study of positronic 
atoms (electronically stable states consisting of a positron 
bound to an atom) 
\cite{mitroy99c,dzuba99,dzuba00a,bromley00a,bromley02a,bromley02b,bromley02d,bromley02e,saito03a,saito03c} 
and also to positron-atom scattering states \cite{bromley03a,novikov04a,gribakin04b}. 
The attractive electron-positron interaction leads to the formation of a 
Ps cluster (i.e. something akin to a positronium atom) in the outer 
valence region of the atom \cite{ryzhikh98e,dzuba99,mitroy02b,saito03a}.

The accurate representation of a Ps cluster using only single particle 
orbitals centered on the nucleus requires the inclusion of orbitals 
with much higher angular momenta than a roughly equivalent electron-only 
calculation \cite{strasburger95,schrader98,mitroy99c,dzuba99}.  
In the most extreme case so far considered, 
namely $e^+$Li, a calculation with $J = 30$ was required before the energy 
had decreased enough to establish binding.  Given that helium is described as
slowly convergent \cite{schwartz62a}, one struggles to find an adjective that 
could characterize the convergence properties of positronic systems!     

The two most important expectation values for positronic systems are
the energy, and the rate for electron-positron annihilation.  The
annihilation rate, which is proportional to the expectation of the
electron-positron $\delta$-function, has the inconvenient property that it 
is even more slowly convergent than the energy with respect to orbital 
angular momentum. One has successive increments decreasing
at high $J$ according to \cite{ottschofski97a,bromley02a,gribakin02a}:
\begin{equation} 
\Delta \Gamma^{J}  =  \langle \Gamma \rangle^J - \langle \Gamma \rangle^{J-1} \approx \frac{A_{\Gamma}}{(J+{\scriptstyle \frac{1}{2}})^2}  \ , \label{pG1} \\   
\end{equation} 
To put this in perspective, it would take a calculation with 
$J \approx 250$ to recover 99$\%$ of the PsH annihilation rate 
\cite{mitroy06a}.   In addition to the slow convergence with $J$,
the $\delta$-function operator also exhibits very slow convergence
with respect to the radial basis \cite{mitroy06a}.  
 
In the present work, large basis CI calculations of the He ground 
state are performed in order to more exactly understand the
convergence of the CI expansion.  Since the properties of the He 
ground state are known to high precision it is a very useful 
laboratory system with which to test methods of extrapolating the 
radial and partial wave expansions to completion.  The insights 
obtained from helium should then be applicable to positronic 
systems and also possibly give additional guidance about how to 
approach purely electronic systems.  Besides looking at the energy, 
the convergence of the CI expansion of the electron-electron 
$\delta$-function expectation value is also studied due to its 
relation with the electron-positron annihilation operator 
(which is also a $\delta$-function).  It should be noted that the 
$\delta$-function operator also appears in the Breit-Pauli 
relativistic correction as the two-body Darwin interaction 
\cite{ottschofski97a,halkier00a}.  The present work builds 
on an earlier investigation that studied the convergence of 
the radial basis in a simplified model of the helium atom 
which only included $l = 0$ orbitals \cite{mitroy06c}.      

\section{The CI method and convergence properties}  

The CI wave function in a single-center basis is  
a linear combination of anti-symmetrised two-electron states 
with the usual Clebsch-Gordan coupling coefficients,   
\begin{eqnarray}
|\Psi;LS \rangle = \sum_{i,j} c_{ij} \: \mathcal{A}_{ij} \:
     & \langle \ell_i m_i \ell_j m_j|L M_L \rangle &
     \langle {\scriptstyle \frac12} \mu_i {\scriptstyle \frac12} \mu_j|S M_S \rangle \nonumber \\  
  \times \: & \phi_i({\bf r}_1) \phi_j({\bf r}_2) & \; .
\label{wvfn} 
\end{eqnarray}
The functions $\phi({\bf r})$ are single electron orbitals 
written as a product of a radial function 
and a spherical harmonic:
\begin{equation}
\phi({\bf r})  =  P(r) Y_{\ell m}({\hat {\bf r}}) \ .         
\label{orbital} 
\end{equation}
All observable quantities can be defined symbolically as  
\begin{equation}
\langle X \rangle^{J} = \sum_{L=0}^{J} \Delta X^{L} \ ,  
\label{XJ1}
\end{equation}
where $\Delta X^{J}$ is the increment to the observable that occurs
when the maximum orbital angular momentum is increased from 
$J - 1$ to $J$, e.g.     
\begin{equation}
\Delta X^{J} = \langle X \rangle^{J} - \langle X \rangle^{J-1} \ .  
\label{XJ3}
\end{equation}
Hence, one can write formally 
\begin{equation}
\langle X \rangle^\infty = \langle X \rangle^{J}  + \sum_{L=J+1}^{\infty} \Delta X^{L} \ .  
\label{XJ2}
\end{equation}

The first term on the right hand side will be determined by explicit computation
while the second term must be estimated.  The problem confronting all single 
center calculations is that part of $\langle X \rangle^\infty$ arises from terms 
with $\ell$-values that are not included in the largest explicit calculation. The 
two expectation values that were investigated were that of the energy   
$\langle E \rangle^\infty$  and the electron-electron $\delta$-function,  
$\langle \delta \rangle^\infty = \langle \delta({\mathbf r}_1-{\mathbf r}_2) \rangle^\infty$. 
For helium, terms with $\ell > 2$ contribute only
0.033$\%$ of the total energy.  For purely electronic systems these 
higher $\ell$ terms make a small (but slowly convergent) correction to the 
total energy and other expectation values.   

The extrapolation schemes used later in this paper have their basis
in the work of Schwartz \cite{schwartz62a}, Hill \cite{hill85a} and 
Kutzelnigg and associates \cite{kutzelnigg92a,ottschofski97a}.
Analytic work indicates that the energy increments are given by 
\begin{equation} 
\Delta E^J = \frac{A_E}{(J+{\scriptstyle \frac{1}{2}})^4} 
   + \frac{B_E}{(J+{\scriptstyle \frac{1}{2}})^{5}}
   + \frac{C_E}{(J+{\scriptstyle \frac{1}{2}})^{6}} + \ldots 
\label{Eseries} 
\end{equation} 
where 
\begin{eqnarray} 
A_E &=& -6\pi^2 \int |\Psi(r,r,0) |^2 r^5 dr = -0.074 226  \label{AE} \\  
B_E &=& -\frac{48 \pi}{5} \int |\Psi(r,r,0) |^2 r^6 dr = -0.030 989  \label{BE} 
\end{eqnarray} 
given a two-body wave function $\Psi(r_1,r_2,({\mathbf r}_1-{\mathbf r}_2))$.
No expressions for $C_E$ have been presented.  At large $J$, one expects
the energy increments to be well described by eq.~(\ref{pE1}).     

For the $\delta$-function one can write 
\begin{eqnarray} 
\Delta \delta^J = \frac{A_{\delta}}{(J+{\scriptstyle \frac{1}{2}})^2} 
   + \frac{B_{\delta}}{(J+{\scriptstyle \frac{1}{2}})^{3}}
   + \frac{C_{\delta}}{(J+{\scriptstyle \frac{1}{2}})^{4}} + \ldots 
\label{dseries} 
\end{eqnarray} 
where $A_{\delta}$ is believed \cite{ottschofski97a} to be   
\begin{equation} 
A_{\delta} = -4\pi \int |\Psi(r,r,0) |^2 r^3 dr = -0.04287   
\label{Ad} 
\end{equation} 
(Ottschofski and Kutzelnigg give a formula similar to this for the 
leading relativistic contribution to the energy of two-electron atoms.  
We have assumed the slow $A_{\delta}/(J+{\scriptstyle \frac{1}{2}})^2$  
convergence is due to the two-electron Darwin term).  It should be noted 
that Gribakin and Ludlow \cite{gribakin02a} have also derived an 
expression equivalent to eq.~(\ref{Ad}) in the context of positron
annihilation.  The numerical value was taken from a variational wave 
function of the He ground state with a basis of 250 explicitly correlated 
gaussians and an energy of -2.9037243752 Hartree.  

\begin{table*}[th]
\caption[]{Results of the present set of 20LTO and 35LTO CI calculations 
of He giving the energy $\langle E \rangle^J$ and
delta-function $\langle \delta \rangle^J$ expectation values as a function of $J$
(all energies are given in Hartree, while $\langle \delta \rangle^J$ is in $a_0^3$).
The total number of electron orbitals is $N_{\rm orb}$ while 
the LTO exponent for $\ell = J$ is listed in the $\lambda$ column.  
The results in the three $\langle E \rangle^{\infty}$ rows use inverse power
series of different length to estimate the  
$J \rightarrow \infty$ extrapolation.
}
\label{Hetab1}
\vspace{0.5cm}
\begin{ruledtabular}
\begin{tabular}{lcccccccc}
     &  \multicolumn{4}{c}{20LTO} &
          \multicolumn{4}{c}{35LTO}  \\ \cline{2-5} \cline{6-9}
 $J$ & $\lambda$ & $N_{\rm orb}$ & $\langle E \rangle^J$ & $\langle \delta \rangle^J$ &  
       $\lambda$ & $N_{\rm orb}$ & $\langle E \rangle^J$ & $\langle \delta \rangle^J$   \\ \hline 
 0 & 4.8 &  20 &  -2.879 028 507  & 0.155 789 346 &  8.6 &  44 & -2.879 028 760 &  0.155 766 769  \\
 1 & 7.8 &  40 &  -2.900 515 873  & 0.128 501 540 & 11.6 &  80 & -2.900 516 228 &  0.128 460 082  \\
 2 & 10.1 &  60 & -2.902 766 378  & 0.120 923 186 & 14.4 & 115 & -2.902 766 823 &  0.120 862 126  \\
 3 & 12.1 &  80 & -2.903 320 527  & 0.117 264 315 & 17.2 & 150 & -2.903 321 045 &  0.117 183 496  \\
 4 & 14.0 & 100 & -2.903 517 973  & 0.115 104 494 & 19.2 & 185 & -2.903 518 552 &  0.115 004 651  \\
 5 & 15.5 & 120 & -2.903 605 022  & 0.113 681 991 & 21.2 & 220 & -2.903 605 654 &  0.113 563 078  \\
 6 & 17.1 & 140 & -2.903 649 142  & 0.112 676 622 & 22.8 & 255 & -2.903 649 820 &  0.112 539 353  \\
 7 & 18.7 & 160 & -2.903 673 821  & 0.111 930 245 & 24.8 & 290 & -2.903 674 539 &  0.111 775 243  \\
 8 & 20.1 & 180 & -2.903 688 677  & 0.111 355 981 & 26.5 & 325 & -2.903 689 430 &  0.111 183 690  \\
 9 & 21.5 & 200 & -2.903 698 142  & 0.110 901 652 & 28.0 & 360 & -2.903 698 926 &  0.110 712 681  \\
10 & 22.9 & 220 & -2.903 704 451  & 0.110 534 160 & 29.5 & 395 & -2.903 705 263 &  0.110 329 155  \\
11 & 24.2 & 240 & -2.903 708 815  & 0.110 231 642 & 31.0 & 430 & -2.903 709 652 &  0.110 011 117  \\
12 & 25.5 & 260 & -2.903 711 927  & 0.109 978 870 & 32.5 & 465 & -2.903 712 786 &  0.109 743 369  \\
\multicolumn{3}{l}{Exact \cite{drake96a}} &  -2.903 724 377 034  &  0.106 345 371  &  &  &   &   \\   
\multicolumn{9}{c}{$\langle E \rangle^{\infty}$ and $\langle \delta \rangle^{\infty}$ extrapolations}   \\   
\multicolumn{3}{l}{Method 1} &  -2.903 723 421  &  0.106 943 &   &   &  -2.903 724 362  &  0.106 527 \\    
\multicolumn{3}{l}{Method 2} &  -2.903 723 252  &  0.107 178 &   &   &  -2.903 724 249  &  0.106 630 \\    
\multicolumn{3}{l}{Method 3} &  -2.903 723 205  &  0.107 334 &   &   &  -2.903 724 240  &  0.106 698 \\ 
\end{tabular}   
\end{ruledtabular}  
\end{table*}  

As can be imagined there have been a number of very large CI calculations
upon the helium ground state that have addressed the convergence issue
\cite{carroll79a,salomonson89b,decleva95a,kutzelnigg92a,jitrik97a,sims02a}.   
These calculations can be roughly divided into two classes, those that 
represented the radial wave function on a grid or used piecewise 
polynomials \cite{carroll79a,salomonson89b,decleva95a}, and those that 
describe the radial wave function as a linear combination of analytic 
basis functions \cite{kutzelnigg92a,jitrik97a,sims02a}.  The first 
systematic calculation was the seminal investigation by Carroll, Silverstone 
and Metzger (CSM) \cite{carroll79a} who used a piece-wise polynomial basis  
to construct a natural orbital expansion.  Besides performing some very 
large calculations they also estimated the completeness limit of their 
radial basis.  The largest explicit calculation by CSM will be termed
the CSM calculation while the extrapolated calculation will be denoted as 
CSM$_{\infty}$.  Despite their importance, these calculations have
been largely superseded by the grid-based calculation of Salomonson
and Oster (SO) \cite{salomonson89b} and the B-spline calculation of 
Decleva, Lisini and Venuti (DLV) \cite{decleva95a}.  The SO calculation  
obtained energies, $\langle E \rangle^J$ accurate to about 10$^{-8}$ 
Hartree by extrapolating the radial basis to the 
variational limit.  This extreme level of accuracy has not been achievable 
with the three calculations that used Slater Type Orbitals (STO) to represent 
the radial wave function \cite{kutzelnigg92a,jitrik97a,sims02a}.  Linear
dependence problems become severe as the basis set is expanded toward
completeness.   Indeed, recourse was made to very high precision (REAL*24) 
arithmetic in the Sims and Hagstrom (SH) calculation \cite{sims02a} which
is the largest calculation of this type so far reported.  

\begin{table*}[th]
\caption[]{Results of the 35LTO$^*$ and 35LTO$^*_{\infty}$ CI calculations of 
He for the $\langle E \rangle^J$ and  $\langle \delta \rangle^J$ expectation 
values as a function of $J$ (all energies are given in Hartree, while 
$\langle \delta \rangle^J$ is in $a_0^3$).
The total number of electron orbitals is $N_{\rm orb}$ while 
the LTO exponent for $\ell = J$ is listed in the $\lambda$ column.  
The results in the three $\langle E \rangle^{\infty}$ rows use inverse power
series of different lengths to estimate the  
$J \rightarrow \infty$ extrapolation.
}
\label{Hetab3}
\vspace{0.5cm}
\begin{ruledtabular}
\begin{tabular}{lcccccc}
 &  &    &  \multicolumn{2}{c}{35LTO$^*$} &  \multicolumn{2}{c}{35LTO$^*_{\infty}$}  \\ \cline{4-5} \cline{6-7}   
 $J$ & $\lambda$ & $N_{\rm orb}$ & $\langle E \rangle^J$ & 
        $\langle \delta \rangle^J$  & $\langle E \rangle^J$ 
        & $\langle \delta \rangle^J$   \\ \hline 
 0 & 4.8 &  35  &  -2.879 028 716  &   0.155 774 273  & -2.879 028 766  & 0.155 763 804  \\ 
 1 & 7.8 &  70  &  -2.900 516 172  &   0.128 472 171  & -2.900 516 246  & 0.128 451 020  \\ 
 2 & 10.1 & 105 &  -2.902 766 757  &   0.120 878 722  & -2.902 766 852  & 0.120 845 876  \\ 
 3 & 12.1 & 140 &  -2.903 320 971  &   0.117 204 759  & -2.903 321 084  & 0.117 159 843  \\ 
 4 & 14.0 & 175 &  -2.903 518 472  &   0.115 030 202  & -2.903 518 601  & 0.114 973 165  \\ 
 5 & 15.5 & 210 &  -2.903 605 568  &   0.113 593 010  & -2.903 605 710  & 0.113 523 543  \\ 
 6 & 17.1 & 245 &  -2.903 649 729  &   0.112 573 360  & -2.903 649 884  & 0.112 491 488  \\ 
 7 & 18.7 & 280 &  -2.903 674 443  &   0.111 813 216  & -2.903 674 609  & 0.111 719 074  \\ 
 8 & 20.1 & 315 &  -2.903 689 330  &   0.111 225 558  & -2.903 689 506  & 0.111 119 165  \\ 
 9 & 21.5 & 350 &  -2.903 698 823  &   0.110 758 273  & -2.903 699 007  & 0.110 639 719  \\ 
10 & 22.9 & 385 &  -2.903 705 157  &   0.110 378 300  & -2.903 705 349  & 0.110 247 727  \\ 
11 & 24.2 & 420 &  -2.903 709 543  &   0.110 063 722  & -2.903 709 741  & 0.109 921 236  \\ 
12 & 25.5 & 455 &  -2.903 712 675  &   0.109 799 333  & -2.903 712 882  & 0.109 645 079   \\ 
\multicolumn{7}{c}{$\langle E \rangle^\infty$ and $ \langle \delta \rangle^{\infty}$ Extrapolations}   \\   
\multicolumn{3}{l}{Method 1} &  -2.903 724 243  &  0.106 881 &  -2.903 724 476  &  0.106 328 \\    
\multicolumn{3}{l}{Method 2} &  -2.903 724 123  &  0.106 757 &  -2.903 724 378  &  0.106 341  \\    
\multicolumn{3}{l}{Method 3} &  -2.903 724 109  &  0.106 847 &  -2.903 724 384 &  0.106 334  \\ 
\end{tabular}   
\end{ruledtabular}  
\end{table*}

\section{The present CI calculations}  

The present calculations use a basis set consisting of Laguerre Type 
Orbitals (LTOs) \cite{shull55a,goldman89a,bromley02a,bromley02b}.  
The LTOs of a given $\ell$ are chosen to have a common exponential 
parameter which means they are automatically orthogonal.  Hence, the  
basis can be expanded toward completeness without causing 
any linear dependence problems.  The CI basis can be characterized by 
the index $J$, the maximum orbital angular momentum of any single electron
orbital included in the expansion of the wave function. It should be
noted that all matrix elements were evaluated using gaussian quadrature 
even though the basis functions have an analytical form \cite{bromley02a}.

\begin{table*}[th]
\caption[]{ Comparison of different CI calculations of the He atom 
ground state energy $\langle E \rangle^J$ as a function of $J$.  To 
aid interpretation, the $\Delta E^J$ energy differences 
are also tabulated.  The energies of the SO$_{\infty}$, CSM$_{\infty}$ 
and 35LTO$^*_{\infty}$ are the estimated energies in an infinite radial 
basis.  The 35LTO$^*_{\infty}$ $\langle E \rangle^J$ and $\Delta E^J$ 
are the smoothed values.  The DLV energies for $J \ge 5$ are obtained by adding the 
$\Delta E^J$ from Table IV of Ref.~\cite{decleva95a} to their estimate 
of $\langle E \rangle^4$.
}
\label{Hetab2}
\vspace{0.5cm}
\begin{ruledtabular}
\begin{tabular}{lcccccc}
$J$ & 35LTO  & 35LTO$^*_{\infty}$ &  SH \cite{sims02a} &   CSM$_{\infty}$ \cite{carroll79a} & SO$_{\infty}$ \cite{salomonson89b} &   DLV \cite{decleva95a}   \\ \hline 
 $\langle E \rangle^0$    &  -2.879 028 760 & -2.879 028 766  & -2.879 028 757 &  -2.879 028 765 &  -2.879 028 77 & -2.879 028 767  \\
 $\langle E \rangle^1$    &  -2.900 516 228 & -2.900 516 246  & -2.900 516 220 &  -2.900 516 220 &  -2.900 516 25 & -2.900 516 245  \\
 $\langle E \rangle^2$    &  -2.902 766 823 & -2.902 766 853  & -2.902 766 805 &  -2.902 766 822 &  -2.902 766 85 & -2.902 766 849  \\
 $\langle E \rangle^3$    &  -2.903 321 045 & -2.903 321 084  & -2.903 321 016 &  -2.903 321 079 &  -2.903 321 09 & -2.903 321 079  \\
 $\langle E \rangle^4$    &  -2.903 518 552 & -2.903 518 601  & -2.903 518 465 &  -2.903 518 598 &  -2.903 518 60 & -2.903 518 600  \\
 $\langle E \rangle^5$    &  -2.903 605 654 & -2.903 605 710  & -2.903 605 515 &  -2.903 605 71  &  -2.903 605 72 & -2.903 605 97 \\
 $\langle E \rangle^6$    &  -2.903 649 820 & -2.903 649 884  & -2.903 649 644 &  -2.903 649 88  &  -2.903 649 89 & -2.903 650 24 \\
 $\langle E \rangle^7$    &  -2.903 674 539 & -2.903 674 609  & -2.903 674 328 &  -2.903 674 59  &  -2.903 674 62 & -2.903 675 01 \\
 $\langle E \rangle^8$    &  -2.903 689 430 & -2.903 689 505  & -2.903 689 193 &  -2.903 689 47  &  -2.903 689 52 & -2.903 689 93 \\
 $\langle E \rangle^9$    &  -2.903 698 926 & -2.903 699 006  & -2.903 698 656 &  -2.903 698 95  &  -2.903 699 02 & -2.903 699 44 \\
 $\langle E \rangle^{10}$ &  -2.903 705 263 & -2.903 705 349  & -2.903 704 974 &  -2.903 705 27  &  -2.903 705 37 & -2.903 705 79 \\
 $\langle E \rangle^{11}$ &  -2.903 709 652 & -2.903 709 742  & -2.903 709 325 &  -2.903 709 64  &                 & -2.903 710 19 \\
 $\langle E \rangle^{12}$ &  -2.903 712 786 & -2.903 712 882  & -2.903 712 433 &                 &                 & -2.903 713 33 \\
\multicolumn{7}{c}{$\Delta E^J $ increments}   \\   
 $\Delta E^1$    & -0.021 487 468 & -0.021 487 480 & -0.021 487 463 &  -0.021 487 455 &  -0.021 487 48 & -0.021 487 478 \\
 $\Delta E^2$    & -0.002 250 594 & -0.002 250 606 & -0.002 250 585 &  -0.002 250 662 &  -0.002 250 61 & -0.002 250 604 \\
 $\Delta E^3$    & -0.000 554 223 & -0.000 554 232 & -0.000 554 211 &  -0.000 554 197 &  -0.000 554 23 & -0.000 554 230 \\
 $\Delta E^4$    & -0.000 197 507 & -0.000 197 516 & -0.000 197 449 &  -0.000 197 519 &  -0.000 197 52 & -0.000 197 521 \\
 $\Delta E^5$    & -0.000 087 102 & -0.000 087 109 & -0.000 087 050 &  -0.000 087 112 &  -0.000 087 11 & -0.000 087 37 \\
 $\Delta E^6$    & -0.000 044 166 & -0.000 044 174 & -0.000 044 129 &  -0.000 044 17  &  -0.000 044 18 & -0.000 044 27 \\
 $\Delta E^7$    & -0.000 024 719 & -0.000 024 725 & -0.000 024 683 &  -0.000 024 71  &  -0.000 024 73 & -0.000 024 77 \\
 $\Delta E^8$    & -0.000 014 891 & -0.000 014 896 & -0.000 014 866 &  -0.000 014 88  &  -0.000 014 90 & -0.000 014 92 \\
 $\Delta E^9$    & -0.000 009 496 & -0.000 009 501 & -0.000 009 463 &  -0.000 009 48  &  -0.000 009 50 & -0.000 009 51 \\
 $\Delta E^{10}$ & -0.000 006 337 & -0.000 006 342 & -0.000 006 318 &  -0.000 006 32  &  -0.000 006 35 & -0.000 006 35 \\
 $\Delta E^{11}$ & -0.000 004 389 & -0.000 004 394 & -0.000 004 351 &  -0.000 004 37  &                & -0.000 004 40 \\
 $\Delta E^{12}$ & -0.000 003 134 & -0.000 003 139 & -0.000 003 108 &                 &                 & -0.000 003 14 \\
\end{tabular}
\end{ruledtabular}
\end{table*}

Three sets of calculations have been performed for the He ground state.  In
the first set, there were 20 LTOs per $\ell$ with the largest calculation
including orbitals up to $\ell = 12$.  The LTO exponents for a given $\ell$ 
were the same and the values of the exponents were optimized in a 
quasi-perturbative fashion.  The exponents for $\ell = 0, 1$ and 2 orbitals
were optimized in a CI calculation with all 60 orbitals.  The exponents 
for $\ell > 2$ were optimized separately for each $\ell$ with CI 
calculations that also included the $\ell = 0, 1, 2$ orbitals.  Once
the exponents were optimized, a sequence of calculations to give the
$\langle E \rangle^J$ and $\langle \delta \rangle^J$ for successive $J$ 
were carried out.  The basis is denoted the 20LTO basis and the results 
of the calculations with this basis are reported in Table \ref{Hetab1}.   

The second set of calculations were much larger.  Here there were 35 LTOs 
per $\ell$ with the exception of $\ell = 0$ and 1 where respectively 44 
and 36 LTOs were used respectively.  The orbital exponents were optimized  
for each $\ell$ in a manner similar to that described above and the
calculations were taken to $J = 12$.  A total of 465 single electron 
orbitals were included in the largest calculation, which required the 
diagonalization of a hamiltonian matrix of dimension 8586.  This 
calculation was an example of a very large explicit calculation.
The basis is denoted the 35LTO basis and the results 
of the calculations with this basis are reported in Table \ref{Hetab1}.   

The idea behind the third calculation was to exploit extrapolation
techniques to estimate the variational limit for each partial wave.
A sequence of calculations with 32, 33, 34 and 35 LTOs per $\ell$ 
was done for a basis that was defined with the same exponential
parameters as the 20LTO calculation.  The number of basis functions   
were varied so that all partial waves had the same basis dimension.
Optimizing the LTO basis 
for the largest radial basis has been shown to result in distortions 
in the convergence pattern with respect to the number of radial
basis functions.  This can be avoided if the basis is optimized in 
a basis that has at least 10 fewer LTOs per $\ell$ than the active 
calculation \cite{mitroy06c}.  The variational limit for the radial 
basis can be estimated by fitting the increments to $\langle E \rangle$ 
and $\langle \delta \rangle$ to the inverse series \cite{mitroy06c}  
\begin{eqnarray}
\Delta E^N &=& \frac{a_E}{N^{7/2}} + \frac{b_E}{N^{8/2}} + \frac{c_E}{N^{9/2}} + \ldots 
\label{Eseries2} \\
\Delta \delta^N &=& \frac{a_{\delta}}{N^{5/2}} + \frac{b_{\delta}}{N^{6/2}} + \frac{c_{\delta}}{N^{7/2}} + \ldots
\label{dseries2}
\end{eqnarray}
It is possible to estimate the $N \to \infty$ limits for the radial basis 
once the $a_E$, $a_{\delta}$, $b_E$, $\ldots$ coefficients have been 
determined.  A two-term series was used for both eqs.~(\ref{Eseries2}) 
and (\ref{dseries2}). It would have been preferable to use three-term 
series but the impact of round-off error rendered this impractical.  
The basis for this set of calculations is denoted the
35LTO$^*$ basis, while the basis including the $N \to \infty$ correction is 
termed the 35LTO$^*_{\infty}$ basis.  The energies and expectation values
for these two calculations are listed in Table \ref{Hetab3}.   

\section{Investigation of the partial wave sequence} 

The validity of these results can be tested by examination of the energy 
increments of large CI calculations of helium.  Besides the present calculations, 
data from a number of previous CI calculations have been used.   

\begin{figure}
\centering
\includegraphics[width=8.5cm,angle=0]{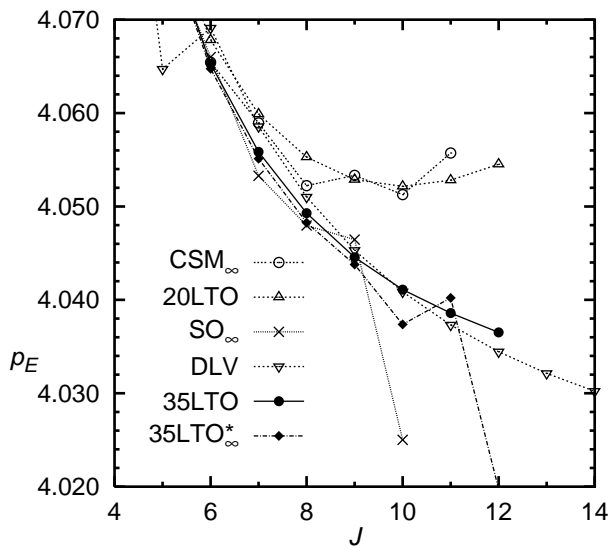}
\caption[]{
The exponents $p_E$ as a function of $J$ for the different 
CI calculations of the He ground state energy as listed in
Table \ref{Hetab2}.  
}
\label{pEHe}
\end{figure}

Table \ref{Hetab2} gives the energies of the present 35LTO and 35LTO$^*_{\infty}$ 
basis sets, along with the SH, CSM$_{\infty}$, SO$_{\infty}$ and DLV calculations.  
These same sets of data are also presented as energy differences between consecutive 
calculations $\Delta E^J$.  The energies of the SH calculation, which used the 
even-tempered STO basis, are consistently the worst, and are $3 \times 10^{-7}$ 
Hartree larger than the 35LTO calculation at $J = 12$.  Even though  
CSM$_{\infty}$ does attempt to achieve the variational limit for each $J$, in 
reality it is only  
about as good as the 35LTO calculation.  Indeed for $J > 7$, the CSM$_{\infty}$ 
values of $\Delta E^J$ were smaller than those of the 35LTO calculation.  The 
present 35LTO$^*_{\infty}$, SO$_{\infty}$ and DLV calculations are in agreement 
to 10$^{-8}$ Hartree (or better) for $J \le 4$.   This is expected, since all 
three calculations are large and use extrapolation techniques to achieve the 
variational limit.  The energy difference between the 35LTO and 35LTO$^*_{\infty}$ 
energies gives an indication of the incompleteness of the 35LTO basis and by 
$J = 12$ the difference is $0.96 \times 10^{-7}$ Hartree.

The good agreement between the 35LTO$^*_{\infty}$, SO$_{\infty}$ and DLV energies 
is not present for $J \ge 5$.  Although the  35LTO$^*_{\infty}$ and SO$_{\infty}$ 
energies generally agree at the level of 10$^{-8}$ Hartree, it is seen the 
DLV $\Delta E^J$ increments are larger than these two other calculations.  For
example, DLV give $\Delta E^5 = 8.737 \times 10^{-5}$ Hartree which is about
$2 \times 10^{-7}$ Hartree larger than the 35LTO$^*_{\infty}$ and SO$_{\infty}$ 
increments.  It has also been noticed that DLV do overstate the accuracy of their
calculation, they assert an accuracy of $7.8 \times 10^{-8}$ Hartree.  However, 
this accuracy is based on a calculation which gives 
$\langle E \rangle^3 = -2.903 319 811$ Hartree (2nd column of Table IV of \cite{decleva95a}), 
and this energy is in error by $1.3 \times 10^{-6}$ Hartree!   

\subsection{Scrutiny of the partial wave increments} 

A useful way to scrutinize the partial wave series is to assume a power 
law decay of the form
\begin{equation}
\Delta X^{J}  \approx  \frac{A_E}{(J+{\scriptstyle \frac{1}{2}})^p}  \ ,     
\label{Xpdef} 
\end{equation}
and determine the value of $p$ for a succession of three $\langle X \rangle^J$ values using 
\begin{equation}
p =   \ln \left(  \frac {\Delta X^{J-1}}{\Delta X^J} \right) \biggl/ 
      \ln \left( \frac{J+{\scriptstyle \frac{1}{2}}}{J-{\scriptstyle \frac{1}{2}}} \right) \ .  
\label{pdef} 
\end{equation}
The exponent derived from the energy increments is $p_E$ while 
the exponent derived from the $\delta$-function increments is 
$p_{\delta}$.  One expects $p_E \to 4$ \cite{schwartz62a} and 
$p_{\delta} \to 2$ as $J \to \infty$ \cite{ottschofski97a,gribakin02a},
in agreement with eqs.~(\ref{Eseries}) and (\ref{dseries}).

The values of $p_E$ for the He energies presented in Table \ref{Hetab2} 
are plotted in Figure \ref{pEHe} as a function of $J$.   One of the 
noticeable features of Figure \ref{pEHe} are the irregularities in 
some of the calculations, e.g. the SO$_{\infty}$, CSM$_{\infty}$ 
and 35LTO$^*_{\infty}$ calculations.  The fluctuations in the present 
35LTO$^*_{\infty}$ curve are due to the impact of round-off error on 
the radial extrapolations.  The determination of the coefficients in 
eq.~(\ref{Eseries2})  involves the subtraction of the energies for 
calculations that differ by a single LTO.  The resulting energy 
differences are very small and therefore are susceptible to the 
essentially random errors resulting from round-off that gradually 
accumulate during the course of the computations.  The irregularities 
in the CSM$_{\infty}$ and SO$_{\infty}$ curves are a consequence 
of the number of digits at which the energies were published 
\cite{carroll79a,salomonson89b}.  Plots of $p_E$ vs $J$ were
examined (but not plotted in Figure \ref{pEHe}) for some 
calculations \cite{jitrik97a,sims02a} that used an STO basis 
set.  These plots of $p_E$ showed much larger fluctuations than 
any of the calculations depicted in Figure \ref{pEHe}.  
    
\begin{figure}
\centering
\includegraphics[width=8.5cm,angle=0]{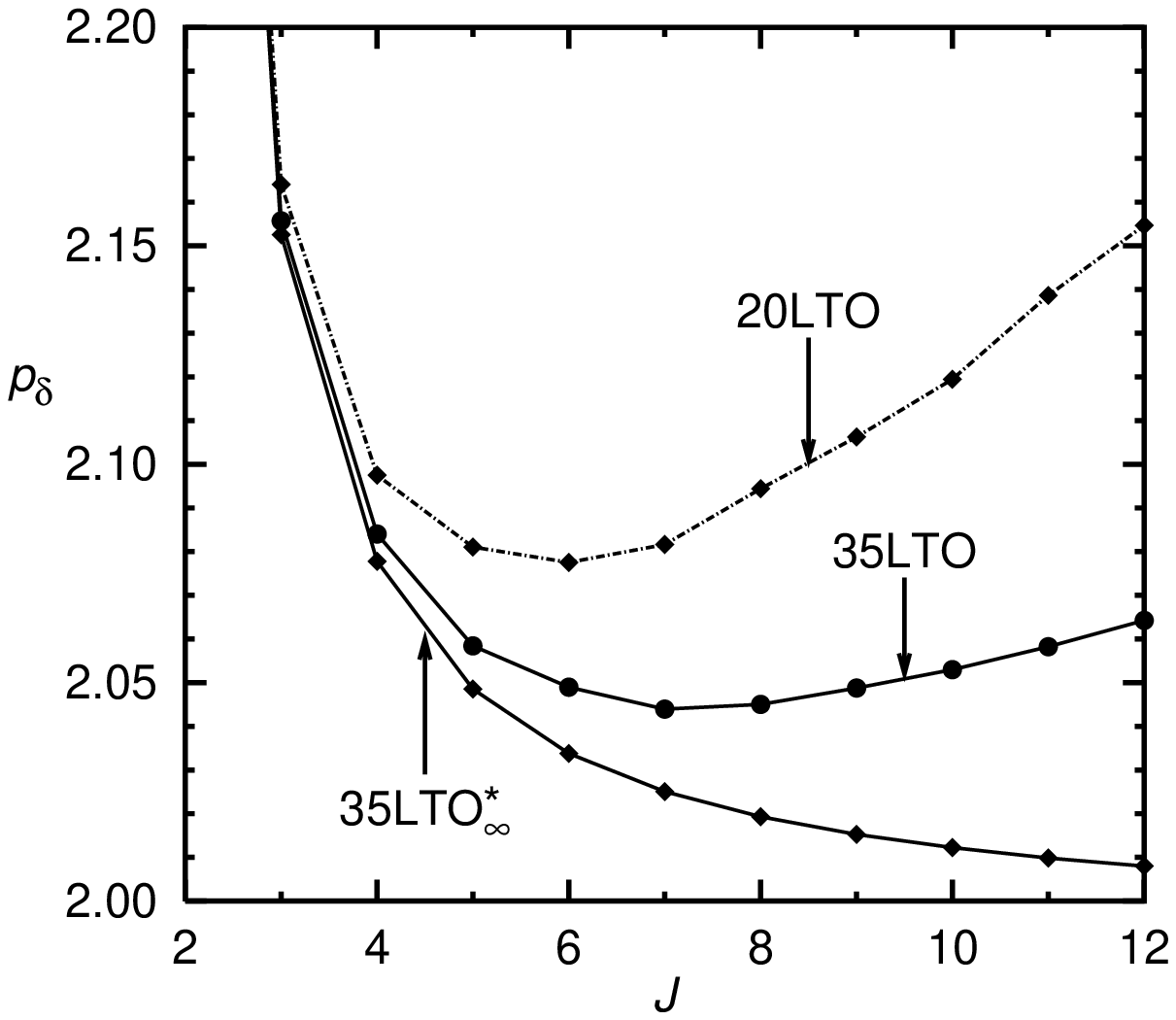}
\caption[]{
The exponents $p_{\delta}$ as a function of $J$ for the 
LTO calculations of the He ground state $\langle \delta \rangle$.
}
\label{pdHe}
\end{figure}

The smaller 20LTO and CSM$_{\infty}$ calculations had plots 
of $p_E$ vs $J$ that tended to level out at $p_E \approx 4.05$.
Indeed, the tendency for 20LTO trajectory to curve up indicates 
that the successive $\Delta E^J$ increments are decreasing too
quickly at the higher $J$ values. The larger SO$_{\infty}$, 
DLV, 35LTO and 35LTO$^*_{\infty}$ calculations have $p_E$ 
versus $J$ trajectories that steadily decrease with increasing 
$J$ and appear to be approaching the expected limit of $p_E = 4$ 
although this is obscured somewhat for the SO$_{\infty}$ and 
35LTO$_{\infty}$ curves.  It will be demonstrated later that the 
behavior of the 20LTO and CSM$_{\infty}$ curves is due to
slower convergence of the radial basis at high $\ell$.      
 
The tendency for $p_E$ to approach the limiting value of 4 from 
above is a consequence of the fact that the $A_E$ and $B_E$ 
coefficients of eq.~(\ref{Eseries}) have the same sign.   The 
coefficients $A_E$ and $B_E$ are derived from second and third 
order perturbation theory respectively \cite{hill85a,ottschofski97a} 
and have the same sign due to repulsive nature of the 
electron-electron interaction.  One surmises that a mixed 
electron-positron system, with its attractive electron-positron
interaction should have $p_E \to 4$ from below, and this is indeed 
the case \cite{bromley02a,bromley02b,bromley02d,bromley02e,mitroy06a}.     

The incremental exponent for the $\delta$-function, $p_{\delta}$, is 
shown in Figure \ref{pdHe} for the 20LTO, 35LTO and 35LTO$^*_{\infty}$ 
basis sets.  It should be noted that the values of $p_{\delta}$ were 
sensitive to the precision of the calculation.  Originally, the 
diagonalization of the Hamiltonian was performed using the Davidson 
algorithm \cite{stathopolous94a}.  However, this method could not give 
$\delta$-function expectation values to better than 8 significant
figures (irrespective of the convergence tolerance for the energy).  
This lead to noticeable fluctuations in the $p_{\delta}$ versus
$J$ plot. The diagonalization was subsequently performed using
the EISPACK libraries, reducing the size of the fluctuations. The 
trajectories of the 20LTO and 35LTO calculations do not appear to 
be approaching the $p_{\delta} \to 2$ limit as $J \to \infty$.   
The 20LTO curve has a $p_{\delta}$ trajectory that diverges from 
2 for $J > 6$ while the 35LTO curve diverges from 2 for $J > 7$. 
The 20LTO basis gives $p_{\delta} = 2.155$ at $J = 12$ while
the 35LTO basis gives $p_{\delta} = 2.064$ at this $J$ value.

However, the plot of $p_{\delta}$ based on the 35LTO$^*_{\infty}$ sequence 
does exhibit the correct qualitative behavior as $J$ increases.  
The $N \to \infty$ corrections have a larger impact on
$\langle \delta \rangle^J$ than on $\langle E \rangle^J$ since
the former converges as $O(N^{-5/2})$ while the latter 
converges as $O(N^{-7/2})$.  The  $J = 12$ value of 
$p_{\delta}$ was 2.008.  

\begin{figure}
\centering
\includegraphics[width=8.5cm,angle=0]{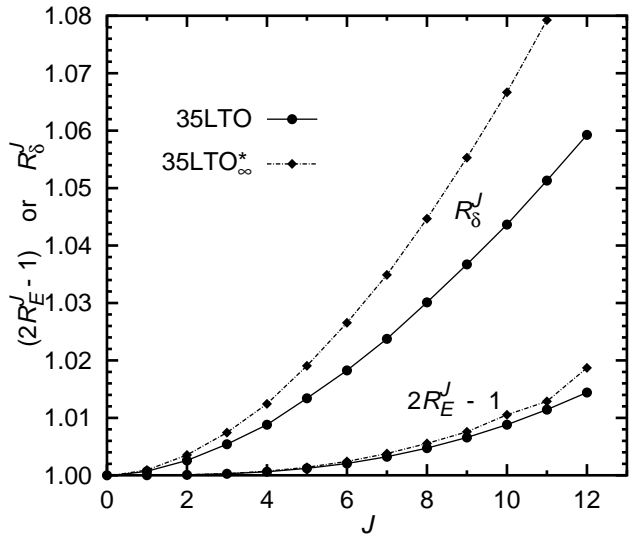} 
\vspace{0.1cm}
\caption[]{
The 35LTO:20LTO and 35LTO$^*_{\infty}$:20LTO ratios of the increments 
to $\langle E \rangle^J$  and $\langle \delta \rangle^J$ (refer to 
eq.~(\ref{ratio})) as a function of $J$ for the calculations of the 
He ground state. 
}
\label{RdE}
\end{figure}

The behavior exhibited in Figures \ref{pEHe} and \ref{pdHe} can
be attributed to the convergence of the radial basis.  A larger 
radial basis is required to predict successive $\Delta E^{J}$
increments as $J$ increases.  The ratios 
\begin{eqnarray}
R^{J}_E = \frac{(\Delta E^{J})_{35}} {(\Delta E^{J})_{20}}    
\label{ratio} 
\end{eqnarray}
gives a measure of the relative impact of the 20LTO, 35LTO and 
35LTO$^*_{\infty}$ basis sets to a $J$ energy increment.  One can 
define a similar ratio,  $R^{J}_{\delta}$, for 
the $\delta$-function $\Delta \delta^{J}$ increments.

The behavior of these ratios versus $J$ are depicted in Figure \ref{RdE}.
Both $R^{J}_{E}$ and $R^{J}_{\delta}$ increase steadily with $J$.  The 
slower convergence of the energy at higher $J$ is also evident in Table I 
of \cite{ottschofski97a} and explicit comment about this point has been 
made previously \cite{kutzelnigg99a}.  The $\langle \delta \rangle^J$ 
expectation value is much more sensitive to the increase in the dimension 
of the radial basis and there was a 9.3$\%$ increase in $\Delta \delta^{12}$ 
between the 20LTO and 35LTO$^*_{\infty}$ calculations.  The corresponding 
increase in the $\Delta E^{12}$ was only $0.8\%$.  This extra sensitivity 
of $\langle \delta \rangle^J$ is something we have noticed in calculations 
of positron annihilation rates in positron-atom systems 
\cite{bromley02b,bromley02d,bromley03a,novikov04a} even though explicit 
mention of this point has not been made.  

\subsection{Smoothing of the 35LTO$^*_{\infty}$ energies} 

It is apparent from Figures \ref{pEHe} and \ref{RdE} that including the 
radial extrapolations has resulted in irregularities appearing in the 
35LTO$^*_{\infty}$ energy sequence.  These irregularities are of order 
10$^{-9}$ Hartree at $J=12$ and should be removed before the 
$J \to \infty$ corrections are determined.  
  
Examination of Figure \ref{RdE} suggested that $(R^J_E-1) \propto J^s$.  Accordingly 
a fit of 
$(R^J_E-1) = \left( \frac{\Delta E^J_{{\rm 35LTO}^*_{\infty}}}{\Delta E^J_{{\rm 20LTO}}}-1\right)$ 
to a $G+H \times J^s$ functional form was performed over the $J \in [5,12]$ 
interval.  An adjustment to the $\langle E \rangle^J$ energy sequence was made
once $G$, $H$ and $s$ were fixed.  The values of $s$ was approximately $s \approx 2.85$.    

The 35LTO$^*_{\infty}$ values of $\langle E \rangle^J$ and $\Delta E^J$ given 
in  Table \ref{Hetab2} are those of the smoothed energy sequence.  The largest change 
to any of the energies was $2 \times 10^{-9}$ Hartree.      

\subsection{Extrapolation of the partial wave series} 
  
\begin{figure}
\centering
\includegraphics[width=8.5cm,angle=0]{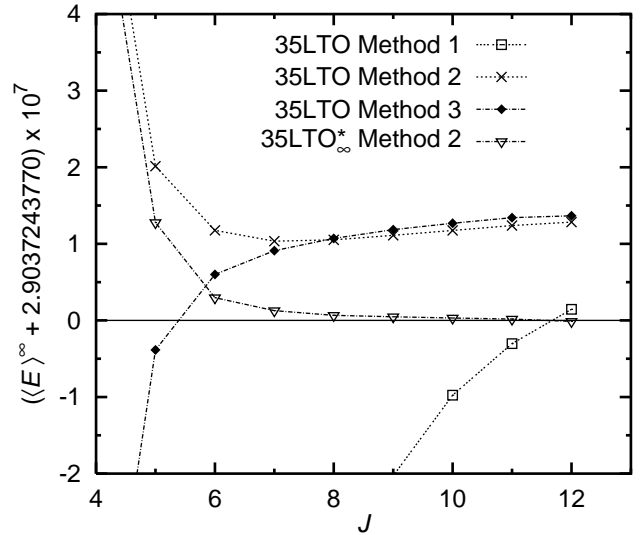}
\vspace{0.1cm}
\caption[]{
The extrapolated $J \rightarrow \infty$ limit for the He ground 
state energy $\langle E \rangle^\infty$ using three different methods
to complete the partial wave series.  The input $\langle E \rangle^J$ 
upon which the extrapolations were based were those of the 35LTO and 
the smoothed 35LTO$^*_{\infty}$ calculations.  The horizontal line 
shows the exact helium energy \cite{drake96a}. 
}
\label{HeEinf}
\end{figure}

One of the major aims of this paper was to determine whether it 
is possible to extract the $J \to \infty$ limit from a finite sequence of 
calculations.  To this end, fits of inverse power series of different 
lengths are made to sequences of $\langle X \rangle^J$ data, and then 
those inverse power series are summed to infinity.

Equations (\ref{Eseries}) and (\ref{dseries}) are the working equations.  
Fits are performed retaining just the leading order term (Method 1), 
the first two terms (Method 2), and the first three terms (Method 3) 
of these series.  The fits of these equations use the minimum information 
necessary.  So, method 1, which only retains the first $A_E$ term of 
eq.~(\ref{Eseries}), only requires two successive values of 
$\langle E \rangle^J$ to determine $A_E$.  Three successive values of 
$\langle X \rangle^J$ are used to determine $A_X$ and $B_X$ when the 
two leading terms of eq.~(\ref{Eseries}) or eq.~(\ref{dseries}) are 
used.  Four successive values of $\langle X \rangle^J$ are used to 
determine $A_X$, $B_X$ and $C_X$ when the three leading terms of 
eq.~(\ref{Eseries}) or eq.~(\ref{dseries}) are used.  The fits to 
determine $A_X$ and/or $B_X$ and/or $C_X$ can be done to different 
sequences of $J$ values as a self-consistency to check that the 
two-term fits to the $J=8,9,10$ or $J=10,11,12$ sets of 
$\langle X\rangle^{J}$ give answers that are numerically
close.   

Once the coefficients of the inverse power series have been determined, 
the $J \to \infty$ contribution is determined by a two-step procedure.
Firstly, the series (\ref{Eseries}) and (\ref{dseries}) 
are summed explicitly up to $J + 200$.  The remainder from $\ge J + 201$ 
is determined using the approximate result:
\begin{equation}
\sum_{L=J+1}^{\infty} \frac{1}{(L+{\scriptstyle \frac{1}{2}})^p } \approx  \frac{1}{(p-1)(J+1)^{p-1} }  \ .
\label{bettertail}
\end{equation}
Eq.~(\ref{bettertail}) can be regarded as an approximation to the
$\int^{\infty}_{J+1} (L+{\scriptstyle \frac{1}{2}})^{-2} \ dL$ integral 
using the mid-point rule.  This approximation is accurate 
to 0.1$\%$ for $p = 2$ and $J = 7$.  

\begin{figure}
\centering
\includegraphics[width=8.5cm,angle=0]{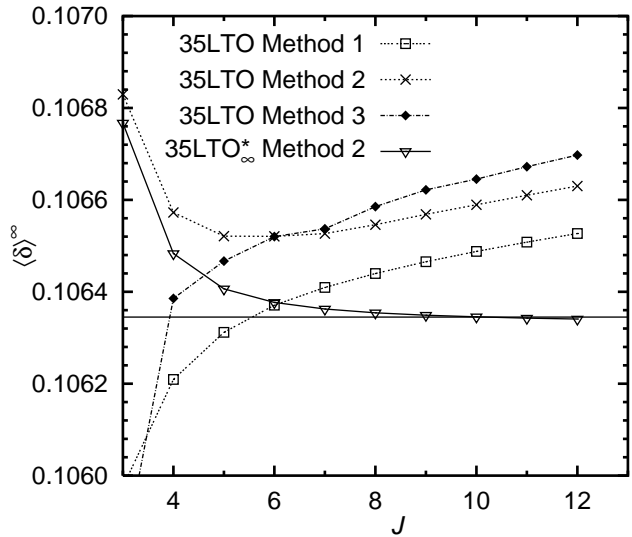}
\vspace{0.1cm}
\caption[]{
The extrapolated $J \rightarrow \infty$ limit for the He ground 
state $\langle \delta \rangle^\infty$ using three different methods
to complete the partial wave series.  
The horizontal line shows the value of Drake \cite{drake96a}. 
}
\label{Hedinf}
\end{figure}

Figures \ref{HeEinf} and \ref{Hedinf} show the behavior of the 
extrapolated $E$ and $\delta$-function for the three 
different extrapolations as a function of $J$.  Tables \ref{Hetab1} 
and \ref{Hetab3} gives estimates of $\langle E \rangle^{\infty}$ and 
$\langle \delta \rangle^{\infty}$ 
using the calculated values at the largest possible $J$ values to 
determine the $J \to \infty$ corrections. 

Figure \ref{HeEinf} shows that the quality of the 35LTO energy 
extrapolation using method 1 is inferior to methods 2 and 3 which give 
$\langle E \rangle^{\infty}$ energies in agreement which each other 
at the 10$^{-9}$ Hartree level for $J \ge 8$.  However, using the
35LTO energies in conjunction with methods 2 and 3 gives  
$\langle E \rangle^{\infty}$ values that are too large by about 
$10^{-7}$ Hartree.  This is a consequence of using a large but 
not quite complete radial basis.   The use of the 35LTO$^*_{\infty}$ 
energies results in an energy limit that is an order of magnitude 
more precise than those of the 35LTO basis.  Using method 2 for
the $J=10, 11, 12$ 35LTO$_{\infty}^*$ energies gave  
$\langle E \rangle^{\infty}$ = -2.903 724 378 Hartree, an energy 
that is in error by 10$^{-9}$ Hartree.  The $J \to \infty$ corrections 
were only made using method 2 since the more sophisticated method 3 is 
more sensitive to the imperfections of the smoothed data sets.      
The smoothed energy sequence is probably not a perfect reproduction
of the actual sequence and there is a tendency for the    
$\langle E \rangle^{\infty}$ limit to be more negative than the 
exact energy.  The method 3 estimate of $\langle E \rangle^{\infty}$,
at $J = 12$, namely -2.903 724 384 Hartree, is about 10$^{-8}$  
more negative than the exact energy.  A similar level of accuracy
was achieved in the earlier SO$_{\infty}$ calculation, their estimate
of the energy in the $J \ to \infty$ limit was -2.903 724 39 Hartree
\cite{salomonson89b}.

The difficulties in obtaining sub 0.1$\%$ accuracy in 
$\langle \delta \rangle^{\infty}$ for the 35LTO sequence 
are readily apparent from Figure \ref{Hedinf}.  As one increases $J$, 
the estimates of $\langle \delta \rangle^{\infty}$ also increase 
and the discrepancy with the accurate value of Drake \cite{drake96a} 
gets larger.  The ultimate accuracy achievable for the 35LTO
basis is between 0.1 and 0.5$\%$.  The apparent superiority
of method 1 for the 35LTO basis arises because errors resulting
from a finite dimension radial basis act to partially cancel 
errors that arise from this least sophisticated $J \to \infty$
extrapolation.  

However, usage of the 35LTO$_{\infty}^*$ sequence permitted a much 
more accurate extrapolation to the $J \to \infty$ limit.  The 
35LTO$^*_{\rm \infty}$ basis gives estimates of 
$\langle \delta \rangle^{\infty}$ that are two orders
of magnitude more precise.  The method 2 extrapolation was only 
$6 \times 10^{-6}$ $a_0^3$ larger than the exact value 
\cite{drake96a}.  While the radial extrapolations did
introduce fluctuations into the $\langle \delta \rangle^J$ values,
the relative size of the individual $\Delta \delta^J$ increments were 
much larger than the $\Delta E^J$ increments and thus did not 
lead to fluctuations in  $\langle \delta \rangle^{\infty}$ 
as long as method 2 was used for the angular extrapolations.  
However, the use of method 3 did result in fluctuations of order 
10$^{-5}$ $a_0^3$ in $\langle \delta \rangle^{\infty}$ so is
not depicted in Figure \ref{Hedinf}.

\subsection{The coefficients of the inverse power series} 

\begin{figure}
\centering
\includegraphics[width=8.5cm,angle=0]{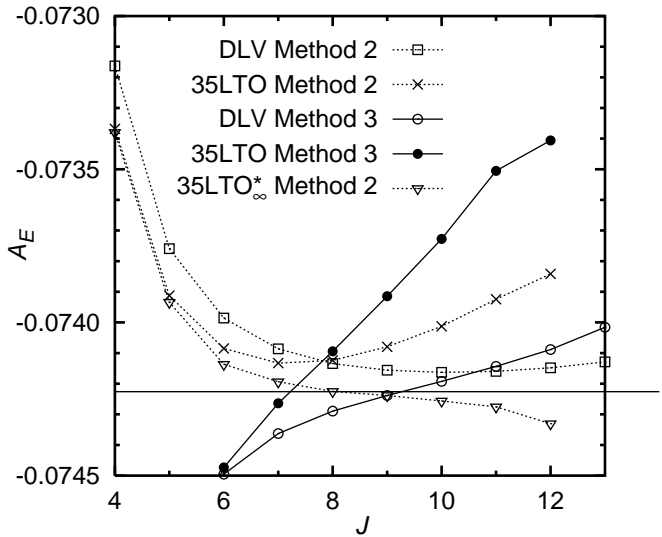}
\vspace{0.1cm}
\caption[]{
The value of $A_E$ as extracted from sequences of $\langle E \rangle^J$ 
data.  The horizontal line shows the value of from eq.~(\ref{AE}), 
namely -0.074226.  
Estimates of $A_E$ from Method 2 are drawn with dashed lines
while estimates of $A_E$ from Method 3 are drawn with solid lines.   
The DLV data analyzed here was solely from their Table IV of Ref.\cite{decleva95a},
since this avoided the discontinuity at $J = 4$ and gave a smooth curve.    
}
\label{AEcoef}
\end{figure}

The coefficients of the asymptotic forms, (\ref{Eseries}) and 
(\ref{dseries}) are known \textit{a-priori} from eqs.~(\ref{AE}), (\ref{BE}) 
and (\ref{Ad}).  Estimates of these parameters are also obtained 
during the fit of the inverse power series to a set of 
$\langle E \rangle^J$ or $\langle \delta \rangle^J$.  An ideal
consistency check would be estimates of $A_E$ and $A_{\delta}$ 
that steadily approached -0.074226 and -0.04287 as $J$ increased
and as the number of terms included in eqs.~(\ref{Eseries}) and 
(\ref{dseries}) increased.  Unfortunately, this has not yet been 
achieved.  The least squares analysis of the CSM$_{\infty}$ 
energies gave $A_E = -0.0740$ and  $B_E = 0.031$ \cite{carroll79a}.  
However, this value of $A_E$ is only achieved when using 
$\langle E \rangle^J$ for $J \in [5,8]$.  The very large calculations 
of DLV reported $A_E = -0.07415$ and $B_E = 0.0317$ \cite{decleva95a}.  
However, a cursory examination of Figure \ref{AEcoef} which
depicts values of $A_E$ obtained from three successive 
$\langle E \rangle^J$ energies demonstrates that  their value of 
$A_E$ is not converging to -0.074226 with increasing $J$.  Applying
the more sophisticated 3-term inverse power series to the DLV
energies leads to an $A_E$ that exhibits a 4$\%$ variation between
$J = 6$ and $J = 13$. 

Figure \ref{AEcoef} also shows the variation in $A_E$ when
fitting the 35LTO and 35LTO$^*_{\infty}$ energy sequences to 
eq.~(\ref{Eseries}).  Fits were performed with both methods 2 and 3 
for the 35LTO energy sets, and only with method 2 (for reasons discussed
earlier) to the 35LTO$^*_{\infty}$ sequence.  The $A_E$ coefficients 
for a given method are computed using the minimum range of $J$ values 
that permitted the unique determination of the coefficients.   

Application of method 2 to the 35LTO data reveals that $A_E$ 
achieves a minimum value of $A_E =$ -0.07413 at $J = 7$ before
increasing at larger $J$.  Application of method 3 results in 
values of $A_E$ that are clearly not approaching the correct 
value.  This should be expected since it has been demonstrated 
that the 35LTO $\Delta E^J$ are increasingly underestimated as 
$J$ increases.   It would therefore be hoped that values of 
$A_E^J$ extracted
from the 35LTO$^*_{\infty}$ would show better convergence to 
the expected limit as $J$ increases.  This expectation has 
only been partly realized, there are indications that $A_E$ 
may be converging to the correct value, but the application of 
smoothing has probably introduced a systematic bias that resulted
in a tendency to overestimate the magnitude of $A_E$.  

\begin{figure}
\centering
\includegraphics[width=8.5cm,angle=0]{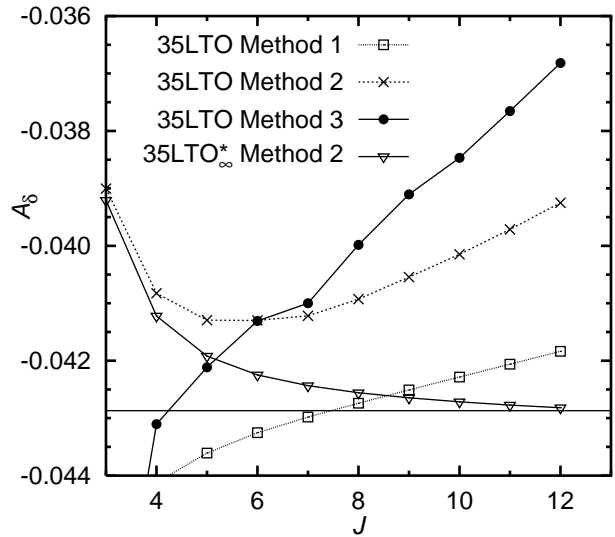}
\vspace{0.1cm}
\caption[]{
The value of $A_{\delta}$ as extracted from sequences of 
$\langle \delta \rangle^J$ data.  The horizontal line shows the value 
of eq.~(\ref{Ad}), namely $A_{\delta} = -0.04287$.   
}
\label{Adcoef}
\end{figure}

Figure \ref{Adcoef} shows the values $A_{\delta}$ as obtained from the
35LTO basis using Methods 1, 2 and 3 as a function of $J$.  None of the 
calculations using the 35LTO basis resulted in an $A_{\delta}$ vs $J$ 
curve that approached the correct value as $J$ increased. This is another 
manifestation of the very slow convergence of $\langle \delta \rangle^J$ 
with respect to the dimension of the radial basis set.  There was a significant 
improvement when $A_{\delta}$ was extracted from the 35LTO$^*_{\infty}$ 
sequence using method 2.  In this case, $A_{\delta}$ does appear to
be converging to the expected value of -0.04287 and at $J = 12$ 
one obtains $A_{\delta} = -0.04282$  

The small irregularities in the 35LTO$^*_{\infty}$ $\langle \delta \rangle^J$ 
sequence resulted in irregularities in $A_{\delta}$ when using 
the more sophisticated and sensitive method 3 fit; so this was not depicted 
in Figure \ref{Adcoef}.   It is also worth noting that $A_{\delta}$ 
was also subject to irregularities of $\pm 2\%$ when the Davidson method
was used to diagonalize the hamiltonian and generate the ground state
wave function.  

\section{Summary and Conclusions}

Results of a set of very large CI calculations of the He ground state 
have been presented.  The largest explicit CI calculation reported 
here with a minimum of 35 LTO's per $\ell$ gave an energy that was 
accurate to $1.2 \times 10^{-5}$ Hartree.  Including energy corrections 
obtained from the 2- and 3-term inverse power series in  $J$ resulted 
in a He ground state energy that was accurate to 
$\approx 1 \times 10^{-7}$ Hartree.  Improved accuracy required the
use of extrapolations in the radial basis set to get an estimate
of the variational limit for $\langle E \rangle^J$.  This permitted
the He ground state energy to be predicted to better than  
10$^{-8}$ Hartree, an improvement of a factor of 1000 over the
largest explicit calculation.  The main impediments to more refined predictions 
of the He ground state are those due to round-off errors.  Estimating the 
coefficients of the inverse power series involves manipulating very 
small energy differences which will be sensitive to round-off errors.  
The fluctuations in the radial extrapolation were about 
$2 \times 10^{-9}$ Hartree at $J = 12$.  While this in itself is not 
that bad, these fluctuations are magnified by an order of magnitude 
when the angular momentum extrapolation is then done.  The impact of the
fluctuations was somewhat mitigated by the introduction of a smoothing
procedure, at the cost of introducing a small systematic error.  

The prediction of the electron-electron $\delta$-function was 
considerably more difficult due to the $O(L+{\scriptstyle \frac{1}{2}})^2$ 
convergence.  In this case, the explicit calculation was accurate to 
3$\%$ at $J = 12$.  Application of the inverse power series (Method 2)
to include higher $J$ contributions improved the accuracy to 0.3$\%$.
The main reason for the low accuracy was the slow convergence 
with respect to the number of radial basis functions.  The relative 
accuracy of successive $\Delta \delta^J$ increments decreases as $J$ 
increases if the number of radial basis functions per $\ell$ is kept 
the same.  Once again, extrapolating the radial basis to the variational 
limit lead to an improved prediction of $\langle \delta \rangle$.  
The best CI estimate of $\langle \delta \rangle = 0.106341$ $a_0^3$ was 
within 0.01$\%$ of the close to exact variational estimate \cite{drake96a}.
The extrapolations of $\langle \delta \rangle$ were less susceptible to
round-off error simply because the $\Delta \delta^J$ increments were
larger.   

While the use of extrapolations did improve the quality of the calculation,
the full potential of the method has not been realized due to round-off 
error.  The radial matrix elements are evaluated with gaussian quadratures
and the achievable precision for the larger calculations is about 10$^{-12}$
Hartree.  This accuracy could be improved by the either the development 
of a convenient analytic form for the electron-electron matrix elements
or the usage of quadruple precision arithmetic.  This would then permit
the use of inverse power series with more terms leading to improved
radial and angular extrapolations.  For example, an accuracy of 10$^{-14}$ 
Hartree was achievable for a CI calculation restricted to $\ell = 0$
orbitals \cite{mitroy06c}.                  
 
These results have implications for the prediction of the annihilation 
rate of positronic atoms from single-center CI type calculations 
\cite{mitroy06a}.  Some sort of extrapolation in $J$ is needed to 
determine the energy and more particularly the annihilation rate.   
One way to minimize the impact of the extrapolation in $J$ is to run the 
calculation to the highest possible angular momentum.  However, the 
high $J$ parts of the annihilation rate will tend to be increasingly 
underestimated as $J$ increases unless accurate estimates of the 
radial variational limit can be made.  Since this can now be achieved for a 
Laguerre basis \cite{mitroy06c}, it eminently conceivable that estimates 
of the annihilation rate at better than $0.1\%$ accuracies will be 
achievable for single-center basis sets.

\section{Acknowledgments}

This work was supported by a research grant from the Australian 
Research Council.  The authors would like to thank Shane Caple 
for providing access to extra computing resources.
One author (MB) would like to thank Prof. Cliff Surko for his,
and his group's, summer hospitality at UCSD while this paper was in preparation.


\end{document}